# The flow of time in the theory of relativity

Mario Bacelar Valente

## Abstract

Dennis Dieks advanced the view that the idea of flow of time is implemented in the theory of relativity. The 'flow' results from the successive happening/becoming of events along the time-like worldline of a material system. This leads to a view of now as local to each worldline. Each past event of the worldline has occurred once as a now-point, and we take there to be an ever-changing present now-point 'marking' the unfolding of a physical system. In Dieks' approach there is no preferred worldline and only along each worldline is there a Newtonian-like linear order between successive now-points. We have a flow of time per worldline. Also there is no global temporal order of the now-points of different worldlines. There is, as much, what Dieks calls a partial order. However Dieks needs for a consistency reason to impose a limitation on the assignment of the now-points along different worldlines. In this work it is made the claim that Dieks' consistency requirement is, in fact, inbuilt in the theory as a spatial relation between physical systems and processes. Furthermore, in this work we will consider (very) particular cases of assignments of now-points restricted by this spatial relation, in which the now-points taken to be simultaneous are not relative to the adopted inertial reference frame.

## 1 Introduction

When we consider any experiment related to the theory of relativity,[1] like the Michelson-Morley experiment (see, e.g., Møller 1955, 26-8), we can always describe it in terms of an intuitive notion of passage or flow of time: light is send through the two arms of the interferometer at a particular moment – the now of the experimenter –, and the process of light propagation takes time to occur, as can be measured by a clock calibrated to the adopted time scale.

However when we consider the theory, in particular focusing on its development in terms of the Minkowski space-time and its application using the Minkowski diagrams, we immediately become puzzled by the implications of the theory regarding our intuitive notion of time, and the passage or flow of time seems difficult or even impossible to 'find' in the conceptual-mathematical structure of the theory (see, e.g., Dorato 2010, 1-2).

Several authors have tried to 'find' in the theory the notions of present/now, becoming, and passage or flow of time, all taken to be closely related (see, e.g., Dieks 1988, Arthur 2006, Savitt 2009). We will address only one proposition made by Dieks (1988, 2006). There is no intention for this work to be conclusive or to give a 'definitive' elaboration of Dieks' approach. The intention is simply to provide, if possible, a further development along Dieks' lines.

[1] Instead of naming Einstein's two theories as special relativity and general relativity, in this work, adopting Fock (1959) terminology, we refer to the theory of relativity and the theory of gravitation. The subject of this work is just the theory of relativity not including Einstein's gravitation theory.

In this work it is made one basic claim. In Dieks' approach it is necessary to consider a consistency requirement regarding the relation between different now-points of different material physical systems. We can consider this requirement to be inbuilt in the theory from the start. It follows from what we can call a spatial relation that exists between the physical systems and processes. For the time being we can characterize a spatial relation as one in which only the 'spatial aspect' is relevant disregarding whatever temporal unfolding these systems and processes might have.

In this work we will also consider particular cases in which, by a further setting of 'initial conditions', the relation between the now-points of different physical systems, in relative (inertial) motion or non-inertial motion, can be determined in a way that does not depend on the particular inertial reference frame adopted. It is important to notice that, in general, this is not possible. The spatial relation between physical systems is not sufficient, by itself, to avoid that the set of now-points of different physical systems taken to be simultaneous is relative to the adopted inertial reference frame.

This paper is organized as follows. In part 2 it is presented Dieks' approach in terms of a now-point per worldline, and mentioned how it enables to maintain a notion of becoming/flow of time in the theory when taking into account the so-called relativity of simultaneity. In part 3 we see how Dieks' consistency requirement is already implicit in the theory, by taking into account that in the theory we consider physical systems and processes that have a spatial relation between them. In part 4 it is addressed the so-called time dilation and several situations in which besides the spatial relation we consider a further stipulation of the initial time phase of different physical systems (in inertial or non-inertial motion). For these particular cases the relation between the now-points of the different physical systems is similar to that of classical mechanics: it does not depend on the adopted inertial reference frame. In the epilogue we return once again to the relativity of simultaneity taking into account the views presented in the work. In the appendix it is addressed the issue of how the flow of time can be seen to be implemented in the Minkowski diagrams.

## 2 Dieks' view on the flow of time in the theory of relativity

Several authors consider that the notion of passage of time is encapsulated in the theory of relativity through the concept of proper time (see, e.g., Dieks 1988; Arthur 2008; Savitt 2009).[2] This view is elaborated in connection with the notion of present or now. The time lapse/passage/flow is related with becoming present, the successive happening of events, or succession of presents (see, e.g., Dieks 1988, 2006; Arthur 2008; Savitt 2009, 2011).[3] In this way the idea that proper time measures/gives the passage of time is entangled with the notion of present or now.

---

[2] The proper time was defined/discovered by Minkowski in his famous 1908 work. By definition the proper time is associated only to material systems to which one always associates time-like worldlines. For this particular case one considers the invariant infinitesimal interval along the worldline at the position of the material system $c^2d\tau^2 = c^2dt^2 - dx^2 - dy^2 - dz^2$. According to Minkowski "the integral $\int d\tau = \tau$ of this magnitude, taken along the worldline from any fixed starting point $P_0$ to the variable end point P, we call the *proper time* of the substantial point at P" (Minkowski 1908, 45). Immediately after this definition, Minkowski applies the concept of proper time as the time gone by a physical system to determine the motion-vector and acceleration-vector of a substantial point (material system).

[3] In this paper it is not presented a detailed analysis of these terms, their possible differences, and philosophical connotations. The only point been made is that the time lapse of a physical system (as measured/given by the proper time) is 'marking' the (local) now of this physical system.

This proposition is made in part as a 'solution' to the difficulty to implement a notion of passage or now in the theory of relativity due to the relativity of simultaneity (see, e.g., Savitt 2013). To see how the problem and its solution arise let us consider first Newton's theory (more exactly classical mechanics). In this case we might identify the flow or transience with the inertial time. The changes, evolution, or motion of physical systems are described in terms of a common temporal parameter – the universal inertial time.[4, 5] This temporal parameter is the same for all physical systems in inertial motion or not. This means, in particular, that all inertial reference frames share the same coordinate (inertial) time.[6] This view can also be expressed in terms of a (Galilian) four dimensional space-time. We can see that the passage of time, as determined by a unique universal time, results in the becoming of a set of events – a hyperplane of simultaneity (the Euclidean space) existing now (see, e.g., DiSalle 2009; Friedman 1983, chapter 3; Savitt 2006, 14; Savitt 2013).

In the case of the theory of relativity it is not possible to consider a hypersurface of events (of the space-time) as simultaneous (i.e. occurring now) for all inertial reference frames.[7] Contrary to the Newtonian case, each inertial reference frame (in relative motion) has a different hyperplane of simultaneous events (i.e. with the same time coordinate), which is a Euclidean space. In this way, there is not a unique set of events corresponding to a unique now shared by all inertial reference frames. This result leads to the relativity of simultaneity.

This situation might imply that there is no place in the theory of relativity for any notion of time flow (or notions taken to be related to it like becoming). For example, according to Gödel (1949), since each observer has a different set of 'nows' (i.e. takes different events to be simultaneous to what he/she considers to be/have been his/her now-point), there is no 'objective' lapse/flow of time. Time would be 'ideal', i.e. a product of consciousness. Gödelian-like views on time in the theory of relativity have 'coalesced' in the so-called block universe view: al events of the Minkowski space-time co-exist tenselessly (see, e.g., Dieks 2006, 168-9; Dorato 2008, 56-9). These views take the time lapse, if it was to have existed in the theory, to have to be given by the coordinate time of the different inertial reference frames. A way out might be to take proper time to give the elapsed time and the becoming present of a physical system (see, e.g. Arthur 2008).

[4] As expressed in the law of inertia, a body in inertial motion takes equal time intervals to travel identical distances in any inertial reference frame: there is a time scale implied in the inertial motion – the inertial time (see, e.g., Torretti 1983, 16-7; DiSalle 1990, 141). Since no inertial body (or inertial reference frame) can be differentiated regarding its motion from other inertial bodies (i.e. all inertial motions are physically equivalent), this means that all inertial bodies share the same inertial time (this can be seen already as a consequence of a restricted form of the principle of relativity expressed in the law of inertia).

[5] Arthur, like Savitt (2011a), defends the view that in the theory of relativity the concept of time bifurcates in two different notions: coordinate time and proper time. We will not address this view in this work. To the purpose of this work it is sufficient to consider that there is a temporal notion in the theory – the proper time, associated to material systems, which does not depend on the adopted inertial reference frame. In the case of an inertial motion (the most important for the views developed in this work), the proper time of a material system is identical to its inertial time; in the case of an accelerated motion, its invariant proper time is calculated from the coordinate time of any adopted inertial reference frame. If there is or there is not a clear-cut distinction between proper time and inertial/coordinate time (i.e. a clear-cut bifurcation of time) does not affect the 'basic' aspect of proper time that is relevant in this work: the flow of time of a physical system is given by its proper time.

[6] According to Torretti there is an implicit assumption regarding the (eventual) synchronization of distant clocks being made to arrive at this result (Torretti 1983, 13). Another possibility is to consider that this result follows from considering Newton's theory in conjugation to his gravitation theory in which the gravitational action-at-a-distance are instantaneous; this 'imposes' an (implicit) synchronization of distant clocks (Brown 2005, 20).

[7] A plane/hyperplane/hypersurface of simultaneity in a particular inertial reference frame is the set of events that have the same time coordinate in this inertial reference frame.

According to Dieks the theory of relatively teaches us that it is not necessary to rely on the idea of a succession of cosmic nows. In his view, "if we want to make sense of becoming we should attempt to interpret it as something purely local" (Dieks, 2006, 157); one must consider the successive happening of physically related events from the perspective of "their own spacetime locations" (Dieks 2006, 157). This points to the centrality of the concept of proper time in the theory of relativity.

As it is well known, only events on the past light cone can affect us (taken our now-point to be the apex of 'our' light cone), and only events in the future light cone can be affected by us (see, e.g., Callahan 2000, 76-7); in relation to us, these events are "unambiguously temporally ordered" (Dieks 2006, 158). Regarding events space-like separated from us, since there is no action-at-a-distance, we cannot have a direct physical interaction with them. How the simultaneity of some of these events to us is defined/determined, has "no influence on the content of our observations" (Dieks 2006, 158), which is taken to be local.[8] According to Dieks,

> we do not need a succession of a definite set of global simultaneity hyperplanes in order to accommodate our experience ... completely different choices of such hyperplanes lead to the same local experiences ... we do not have to bother about global simultaneity at all. If we decided to scrap the term 'simultaneity' from our theoretical vocabulary, no problem would arise for doing justice to our observations. (Dieks 2006, 160)

The importance of the relativity of simultaneity would not be directly related to questions regarding the time ordering of distant events, but in pointing to the crucial aspect that the temporal experience is local.

Dieks proposes to reformulate the idea of flow of time based on the concept of proper time. According to Dieks "only time *along worldlines* (proper time) has an immediate and absolute significance as an ordering parameter of physical processes" (Dieks, 1988, 456). However since there is not in the theory of relativity any preferred worldline (or associated reference frame; as it is the case of Lorentz's electron theory with its preferred ether based reference frame), there is no way to single out a particular worldline and its private now. Accordingly, "it is not appropriate to define one universal 'now'; instead, we have to assign a now-point to every single worldline" (Dieks 1988, 458). The flow of time is contained in the theory if we consider that "each point of the worldline has to occur once as 'now-point'" (Dieks 1988, 458).[9] Dieks' view is that, contrary to earlier views on time, the relativistic framework leads to a generalization of the universal flow of time, which Dieks refers to as flow of time per worldline (Dieks 1988, 459). This approach leads to a 'many-fingered' view of time; each worldline has

[8] This does not mean that we cannot have a local experience related to an event with a space-like separation from us; e.g., we can receive light emitted in/with a particular event that will later arrive at what will be our now-point.

[9] There is a 'variant' to this approach in which the notion of local present/now along a worldline is not restricted to a now-point. The motivation for this proposition seems to be outside physics. While Dieks regards that an (human) observation can be represented by a point-event (Dieks 2006, 3), the notion of psychological present seems to point, according to some, to a physical representation of the now as extended. The events that are part of our now are the ones with which we can interact during a short interval of our proper time. Instead of a now-point, the present is a region of space-time comprised between the future light cone of the 'beginning' $e_1$ and the past light cone of the 'end' $e_2$ of our extended present (see, e.g., Arthur 2006; Savitt 2009; Dorato 2011). It is beyond the scope of this work to address the proposition of an extended now.

its own now-point. However, this view might lead to a consistency problem. According to Dieks

> an arbitrary assignment of now-points to the worldlines will not do, however, for the following reason. The idea of a flow of time combined with the ontological definiteness of present and past requires that everything that is in the past lightcone of an event that is ontologically defined is also ontologically define. This leads to the demand that no now-point should lie in the interior of the conjunction of the past lightcones of the other now-points. (Dieks 1988, 458)

To see this approach at work, let us consider an example given by Petkov (2009). We have two observers A and B in relative motion;[10] we can regard the observers to be located at the origin of two inertial reference frames and carrying a clock each. Let $C_1$ and $C_2$ be two clocks from A's inertial reference frame (at some location -d and +d). As it is usually done, we consider that A and B meet at event M (i.e. that there is a moment in which the origins of the references frames coincide) and that they set the clocks to $t_A = t_B = 5$ (see Figure 1, which gives a schematic representation of the worldliness of A, B, $C_1$, and $C_2$, and the planes of simultaneity of A and B).[11]

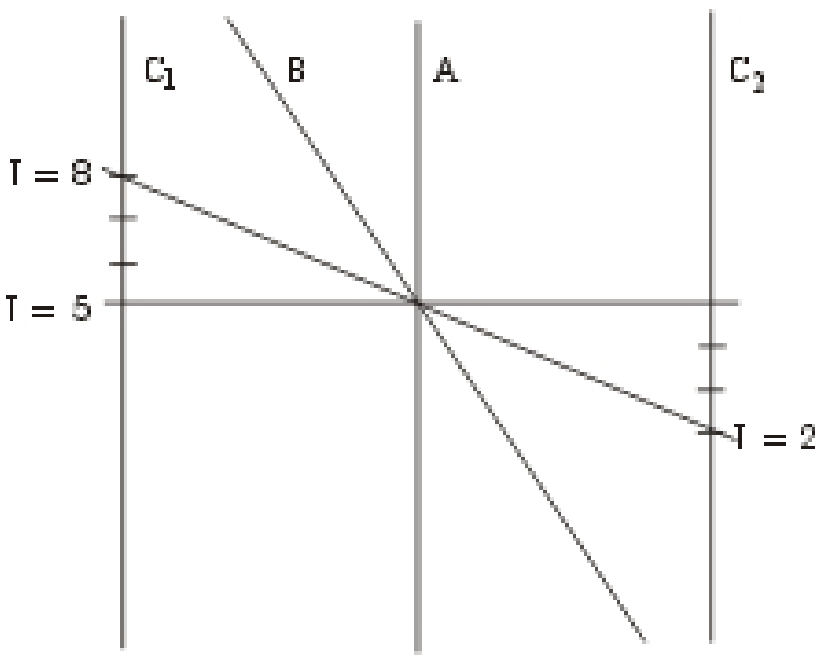

Figure 1

According to Petkov, when identifying the 'now' to the clock's proper time, "to observer A, both clocks exist at the 5th second of the coordinate time measured in A's reference frame" (Petkov 2009, 129); in this way,

> as in an *inertial* reference frame the coordinate (global) time coincides with the proper times of all objects at rest in that frame, it follows that A comes to the conclusion that $C_1$ and $C_2$ both exist at the 5th second of their proper times. (Petkov 2009, 129)[12]

[10] As it is a current practice in some of the physics literature we will adopt the notion of 'observer' as an agent in inertial motion making spatial and time intervals measurements. The agent sets an inertial reference frame – simply by stipulating the phase of a 'master' clock – considering herself to be located at the origin of this 'frame' (see, e.g., Bondi 1965, 71-92; Bohm 1965, 131-45; Ludvigsen 1999, 12-7).

[11] To simplify, in this work we will adopt a unit of time such that the speed of light is taken to be equal to one. This will enable in diagrams to represent the paths (worldlines) of light signals/pulses always by straight lines having an angle of 45º relative to the space and time axes (see, e.g., Bohm 1965, 132; Wheeler and Taylor 1963, 17-8). Also, in this work it will be considered just one spatial direction identified with the letter x.

[12] Petkov identifies the numerical value of the coordinate time with the proper time in the case of an inertial motion. However, even if accepting Arthur's clear-cut bifurcation of time, we must grant to Petkov that when in inertial motion a physical system's inertial time – which is numerically equal in this

The problem with this approach is that due to the relativity of simultaneity, “what is simultaneous for A, however, is not simultaneous for B” (Petkov 2009, 129). When considering B’s plane of simultaneity we see that

> what is simultaneous for B at the $5^{th}$ second of B’s time (when B meets A at M) is clock $C_1$ existing at the $8^{th}$ second of its proper time and clock $C_2$ existing at the $2^{th}$ second of its proper time. Therefore, for B the moment ‘now’ of the proper time of $C_1$ is the $8^{th}$ second, whereas the present moment of $C_2$ is the $2^{th}$ second of its proper time. (Petkov 2009, 129)

As a result, “when A and B meet at M, they will disagree on which is the present moment of each of the clocks” (Petkov 2009, 129). If we accept both hyperplanes of simultaneity as giving the common/global now of all physical systems, we would face the paradoxical situation that $C_1$ and $C_2$ would exist at once in different moments of their proper time.

But with Dieks the now is not defined in terms of simultaneity. The now/present is local to each worldline As Dieks mentions, “we do not need a succession of a definite set of global simultaneity hyperplanes in order to accommodate out experience” (Dieks 2006, 4). The fact that A and B take different elapsed times of $C_1$ and $C_2$ as now for them is not problematic regarding the now-points of $C_1$ and $C_2$ (whatever these might be), if these events are occurring or have already occurred to $C_1$ and $C_2$, i.e. if they are present or past events in the ordered temporal relation within each worldline of each physical system $C_1$ and $C_2$. What is necessary, regarding the relation between the now-points of A, B, $C_1$, and $C_2$, is that the assignment of a now-point to each worldline is made in a way that “no now-point should lie in the interior of the conjugation of the past lightcones of the other now-points” (Dieks 1988, 458). In this approach it seems it might be possible to make compatible a local view on passage and the inexistence of a global now. However there seems to be no physical relation between the different now-points. As it is, by now, in Dieks’ approach, we must regard the now-points as simply set by hand as initial conditions in case-by-case applications.

That is not totally the case. As we will just see, there is in fact a need to set at least the initial relation between the now-points of two inertial observers. But this is constrained and in fact made possible by the fact that these material physical systems are taken from the start, in the development of the theory, to have a particular relation between them, which turn out to be a spatial relation.

## 3 Dieks’ consistency requirement as implicitly determined in the theory from the spatial co-existence of physical systems and processes

Let us consider an inertial reference frame made of a set of inertial bodies – to simplify, clocks – at relative rest. Adopting Dieks’ view of a now-point per worldline we might ask about the passage of time of the clocks of the inertial reference frame and the relation between their respective now-points. Contrary to Einstein’s views we do not need an explicit implementation of a synchronization procedure using the propagation of light to establish the simultaneity between distant clocks (see, e.g., Einstein 1910, 126-7; Darrigol 2005). It is sufficient to choose the phase of a ‘master’ clock and to take into case to its proper time – gives, because of this equality, the time lapse for this physical system.

account the constancy of the one-way speed of light.[13] This already sets a coordinate time associated to the inertial master clock.[14] Being at rest in relation to the master clock the phases of the other clocks are already implicitly determined, being realized, e.g., when an actual light pulse emitted from the master clock reaches another clock.

There is a possible oversimplification being made in this approach. If we think in terms of just one inertial body and its worldline (e.g. the observer with her master clock), then by choosing a particular temporal phase the coordinate time in empty space would be fixed. In this case, we would be taking for granted that there is already a meaningful notion of metrical spatial distance associated to 'empty space'. To avoid thinking in terms of just one material body in empty space we might resort to Einstein's notion of body of reference. In the context of the theory of relativity, Einstein avoids speaking of space (mathematical Euclidean space) in abstract. The (Euclidean) space of reference is thought in terms of the space of/associated to an extended material body – the body of reference, taken to be in inertial motion (see, e.g., Einstein 1955; Paty 1992, 24-5).

We can think of our body of reference, as an inertial reference frame, enabling spatial and temporal determinations, constituted, e.g., by 'elementary' physical systems – rods and clocks. In particular let us consider a measuring rod A (adopted as our unit of length). According to Einstein, let us

> [bring] bodies B, C, . . . up to body A; we say that we continue body A. We can continue body A in such a way that it comes into contact with any other body, X. The ensemble of all continuations of body A we can designate as the 'space of the body A'. (Einstein 1955,6)

All these material continuations of the body A, constituting the body of reference, are spatially present to each other. The master clock and the other clocks can be seen as having a metrical spatial relation determined in the context of this ever-present body of reference.

It terms of Dieks' approach we can see the different inertial bodies (clocks) at relative rest some distance apart in terms of worldlines. Thinking of our master clock as located at the surface of an extended material body (in inertial motion), it seems reasonable to consider that the other clocks at rest in relation to the extended inertial body have their now-points in the elsewhere of the master clock. This is so because, in this case, we expect the interval between the now-points to be space-like.[15] This would imply that, independently of the precise now-point each one might have, they are already in accordance with the consistency requirement mentioned by Dieks regarding the relations be-tween now-points. This, however, does not seem to assign in a unequivocal way the now-points of the clocks of the inertial reference frame.

[13] This implies taking the so-called one-way speed of light to be the same in all directions. There are authors that take this to be a conventional choice (see, e.g., Anderson et al. 1998, 16).

[14] In the theory of relativity the definition of the coordinate time of an inertial body or inertial reference frame from the inertial time can be implemented in a unique way, because we do not rely only on the law of inertia, which determines a unique inertial time but not a unique coordinate time (Torretti 1983, 53). There is besides the law of inertia of material bodies what we might call the law of inertia of the electromagnetic field (Fock 1959, 8; Torretti 1983, 55): light traverses equal distances in equal time intervals with the same speed in all directions. By taking into account light's inertial properties, the coordinate time associated to an inertial body or inertial reference frame is uniquely determined from the inertial time simply by choosing a 'master' clock associated to the inertial body (or inertial reference frame) and setting its 'initial' phase. If a light pulse is send from the position x and arrives at the origin when the master clock reads t, then we know that the event "sending the light pulse" occurred t - x/c seconds ago in terms of the coordinate time of the inertial reference frame.

[15] Considering events $e_1$ and $e_2$ that have a space-like separation, i.e. for which $\Delta x^2 > c^2 \Delta t^2$, one says that $e_2$ is in the elsewhere of $e_1$ (see, e.g. Bohm 1965, 146-154; Schutz 1985, 10-15).

There is however an element that is not yet explicitly taken into account. The inertial bodies at relative rest can, e.g., exchange light. We do not need to assign by hand the now-point of each inertial body at relative rest so that we can ascertain that a light signal send by one of the inertial bodies will reach another one.[16] If we imagine the clocks of the inertial reference frame as inertial worldlines with different now-points previous to defining an inertial reference frame, and from a particular now-point of a clock is emitted a light signal to another clock (in its elsewhere), the second clock's now-point is fixed due to its spatial relation (e.g. the belonging to the extended inertial reference frame) to the first clock, otherwise it would not receive the light. This is an important point since it shows, contrary to Dieks' consistency requirement, that the relation between the now-points of different clocks of an inertial reference frame is at least partially determined by something more than the consistency requirement.

The key aspect to be taken into account is that we are considering from the start physical systems that, e.g., can be brought together side-by-side or moved around in relation to each other. We are considering what we might refer to as physical systems that have a spatial relation between them. They co-exist as spatial things: things that are spatially located in relation to each other (independently of their local unfolding/passage/change).

When we start to speak in terms of 'spatially located', 'spatial thing' or 'spatial relations' it gives the impression that there is something vague and 'unscientific' in this terminology. This does not have to be the case. According to Dieks,

> 'being something spatial' is a quality whose content is not fixed by saying that it belongs to elements possessing [e.g.] the interrelations of the points of the Euclidean plane … to fix the reference to spatial thing something additional must be invoked. A natural move to make is to embed ourselves in the network of relations, and to identify some of the experiential relations between ourselves and the world around us as spatial. (Dieks 2006, 171-2)[17]

An example of a 'natural network' of relations involving our experience in the world is given by practical implementations of the notion of inertial reference frame. Let us consider an extended body like the Earth and satellites orbiting it. In practice one uses a network of satellites to 'calculate' an inertial reference frame (see, e.g., Barbour 1989, 665-6). The satellites are spatially located in relation to each other, to the Earth, and to us; we all co-exist as spatial things.

The realization that physical systems are spatial co-existent makes possible for the particular case of physical systems belonging to/constituting an inertial reference frame to settle/determine a coordinate time shared by all of them. That is, because these physical systems are spatially co-existent we can determine a shared coordinate time in all the inertial reference frame built from the identical inertial time of the physical systems, e.g. by exchanging light between them. Saying this in other words: in this case, the relation between the now-points of the physical systems is fixed; they all share the same temporal parameter.

---

[16] We can imagine e.g. two observers at the surface of an extended body (in inertial motion) some distance apart and exchanging light signals.

[17] Dieks makes these remarks in a context different from that developed in the present paper, in relation to a tentative 'grasping' of the notion of 'temporality'.

It might seem obvious that material bodies can exchange light between them or even be side-by-side at relative rest or momentarily when, e.g., in relative motion. However, when starting from a view in terms of worldlines, each one with its local now-point this result is not self-evident. For example, the possibility that time-like worldlines might cross or that time-like worldlines can have their now-points connected by light-like worldlines (i.e. the worldlines of light signals/pulses) is not 'covered' by Dieks' consistency requirement regarding the relation between the now-points of different time-like worldlines. This follows when taking into account that the theory is built by considering physical systems (like material bodies) and physical processes (like the propagation of light) that co-exist spatially. Dieks' consistency requirement then follows from this spatial co-existence; i.e. for physical systems that are spatially co-existent it follows that the relation between their now-points is such that no now-point of a physical system lies in the interior of the conjugation of the past light cones of the other physical systems.

We can imagine, for this particular case, that our inertial bodies are part of/exist in a three-dimensional spatial 'manifold' whose spatial metric relation is not determined independently of this set of inertial bodies at relative rest chosen to constitute an inertial reference frame. Each of these physical systems immersed in the spatial 'manifold' has a temporal unfolding given by its proper time. In the case of any inertial reference frame the relation between the unfolding/becoming of its 'members' is fixed and can be described in terms of a common temporal parameter – the coordinate time.

## 4 A new look into Dieks' now-point per worldline view taking into account the spatial co-existence of physical systems

As mentioned, all material bodies in inertial motion share the inertial time. This implies, in particular, that if we consider a clock in relative motion in relation to an inertial reference frame, it has the same rate as all the clocks of the inertial reference frame. However, as it is well known, from the perspective of the inertial reference frame the clock in relative motion has a smaller rate (this is the so-called time dilation). But this is also the case when measuring with the clock in relative motion the rate of any clock of the inertial reference frame; the clocks of the inertial reference frame will appear to run at a slower rate. This might give the impression that the time dilation is not real. In fact, the relative retardation of clocks moving inertially in relation to each other has nothing to do with the rate of the clocks as such, otherwise the principle of relativity would not be valid (see, e.g., Bohm 1965, 131-40; Smith 1965, 57; solution to problem 30 in chapter 1 of Wheeler and Taylor 1963).

Let us consider a clock A located side-by-side with an identical clock O of the inertial reference frame. At the moment that clock O gives/reads zero, clock A is boosted to a state of relative motion with an initial phase of zero (see, e.g., Brown 2005, 30). It is evident that in Dieks' terms their initial now-points coincide. This is not in disagreement with Dieks' consistency requirement regarding the relation between different now-points. In terms of the view presented here both clocks co-exist spatially, and this implies, in particular, that when side-by-side they have the same now-point. Both clocks send light pulses to the other $T_0$ seconds apart. The successive light pulses arrive at the other clock $T = kT_0$ seconds apart, where $k = \mathrm{sqrt}((1 + \upsilon/c)/(1 - \upsilon/c))$. This is the relativistic Doppler effect (see figure 2). Let us determine the rate of clock A as measured by clock O. When clock A measures a time interval of $T = kT_0$, it is located in relation to clock O in a position corresponding to a time t as measured by clock O. Calculating the time reading t measured by the clock O that corresponds to the time reading T measured by the clock A, we have t = T/ sqrt(1 - $\upsilon^2/c^2$). If we calculate in exactly the same way the rate of clock O as measured by clock A we arrive at the same result (see, e.g., Bohm 1965, 134-140). The measurement of the time gone by clock A (O) made by clock O (A) as predicted by the theory is in agreement with experimental results (Zhang 1997, 175-200); in this sense we take the time dilation to be 'real'.

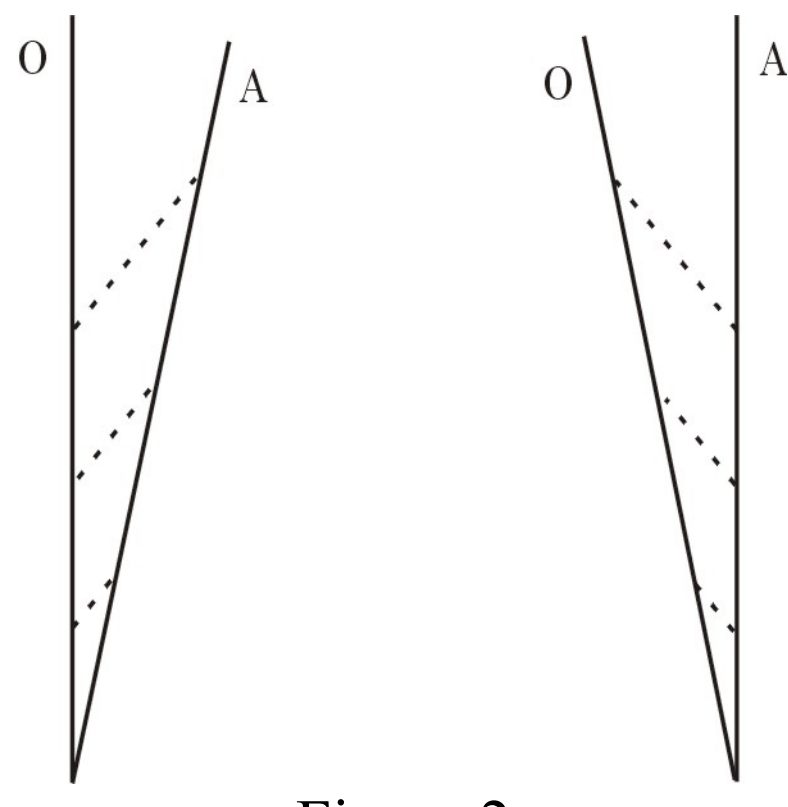

Figure 2

When analyzing the time dilation in terms of space-time diagrams (including the 'axes' of two inertial reference frames, but choosing one of the inertial reference frames as the 'rest' frame; see figure 3) the situation might seem incompatible with the possibility of having the same temporal value of the becoming (i.e. both now-points having the same value of the time parameter). When observer A reads, e.g., 10s in her master clock she, later, upon receiving light emitted by observer B at his now-point corresponding to, e.g., t' = 8s, considers that this is the now-point of B that was now with her t = 10s. In the same way observer B takes the now-point of A with t = 8s to have been now with his t' = 10s. However if we take the diagram at 'face value' it might seem that A's t = 10s is to A simultaneous with B's t' = 8s (corresponding to the S plane) *and* B's t' = 10s is to B simultaneous with A's t = 8s (corresponding to the S' plane). This might give the impression that the diagram (and the Minkowski space-time) is incompatible with any notion of flow of time or passage. It seems that we have two simultaneity planes (S and S') given 'at once' in a fixed diagrammatic representation of the four-dimensional manifold (see, e.g., Petkov 2005).

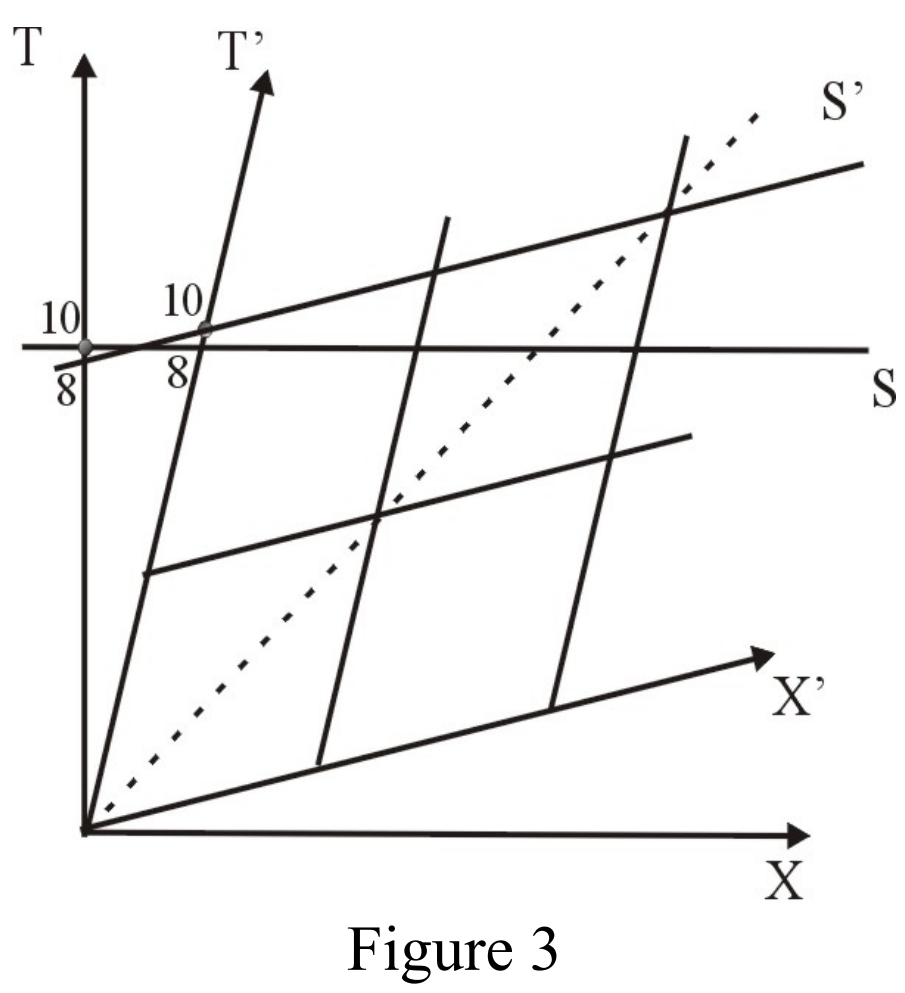

Figure 3

However we must recall that if the now of a physical system is given by its proper time and the principle of relativity is valid, it follows that both observer A and B must be taken to read 10s at the 'same time', i.e. as their corresponding now-points: since the clocks when side-by-side had the same now-point and both have a local flow of time according to inertial time, their current space-time now-points have the same temporal value. We can conclude that, even if they are in the elsewhere of each other, if one of the clocks has gone by t seconds the other clock in relative motion has gone also by t seconds. This follows from the principle of relativity.

Returning to our image of a three-dimensional spatial 'manifold', if we imagine being outside the spatial 'manifold' seeing instantaneously the evolution of each physical system, we would see both master clocks reading 10s. However our experiments and observations must be described as 'internal' to the spatial 'manifold'. As Dieks calls the attention to, the content of our observations is local. The temporal relation between distant events can only be reconstructed through, e.g., the exchange of light; and this introduces a 'distortion' in the determination of the now-points corresponding to an observer's now-point. Both A and B are equivalent physical systems in inertial motion with identical clocks; when side-by-side they set the phase of their clocks to zero. If one goes by t seconds then the other also goes by t seconds. If observer A wants to determine the temporal values t' and t corresponding to the physical event "reflection by B of light send by A", she can determine t by sending a light pulse to B at her time reading $t_0$, which is reflected by B and arrives back at A at her time reading $t_1$.

We can imagine that the light pulse send back by B 'carries' the information of B's proper time reading t' at the moment that light is reflected by him. This means that when the light returns to A she has access to B's time reading. Then A calculates, by taking into account that the speed of light is the same in both directions and independent of her or B's (relative) motion, the event along her worldline that corresponds to that particular event in B's worldline. To A the time reading of observer B when he sends back the light pulse is $t = t_0 + (t_1 - t_0)/2$ (see, e.g., Bohm 1965, 143).

The relation between the measured value t' and the partially measured and partially calculated value t agrees, within the scope of application of the theory, with the theoretical prediction: as we have just seen, B's time reading (as determined by his proper time) when the light is reflected by him is equal to $kt_0$, i.e. the temporal value of the now-point corresponding to the reflection of the light pulse is according to B $t' = t \text{ sqrt}(1-\upsilon^2/c^2)$.

This 'experimental procedure' (or any other) to measure according to A the time reading of B enables a reconstruction at A's 'current' now (given by $t_1$) of the temporal relation between past events of A and B that had a space-like interval between them.

When A and B determine, by exchanging light, the events of the other corresponding to their current/past now-points, they arrive at the result that there is a time dilation. It gives the impression that when A is e.g. in her now-point with t = 10s, B is so to speak in the past in his now-point corresponding to t' = 8s. The same happens to B, which when reading 10s in his master clock will, *later by calculation/measurement*, determine that A's corresponding now-point *was* in his past, since B considers that A's now-point corresponded to her t = 8s. However that is not the case; A and B can calculate away the relativistic effect simply by recalling that both have the same proper time reading according to the principle of relativity: when A's clock is reading 10 seconds B's clocks is also reading 10 seconds.

This result, as such, only applies to the particular case of these two inertial bodies that besides being in spatial co-existence (a necessary prerequisite) defined their respective coordinate times by setting their phases to the same value when side-by-side. If we consider another inertial body C in relative motion that was not side-by-side with A and B the situation is more intricate. We start with the knowledge that C co-exists with A and B; this means that even if due to its particular state of inertial motion body C might not ever be momentarily side-by-side with A or B it can nevertheless exchange light with them. This restricts the 'position' of its now-point in relation to the now-points of A and B. Let us consider that when t = t' = 0 light is emitted from A and B and that it arrives at C when A's clock is reading T (A can determine this value calculating it from the time 2T that light takes to arrive back at A; see figure 4). We know that B attributes a different time of arrival according to his clock, let us call it T'.

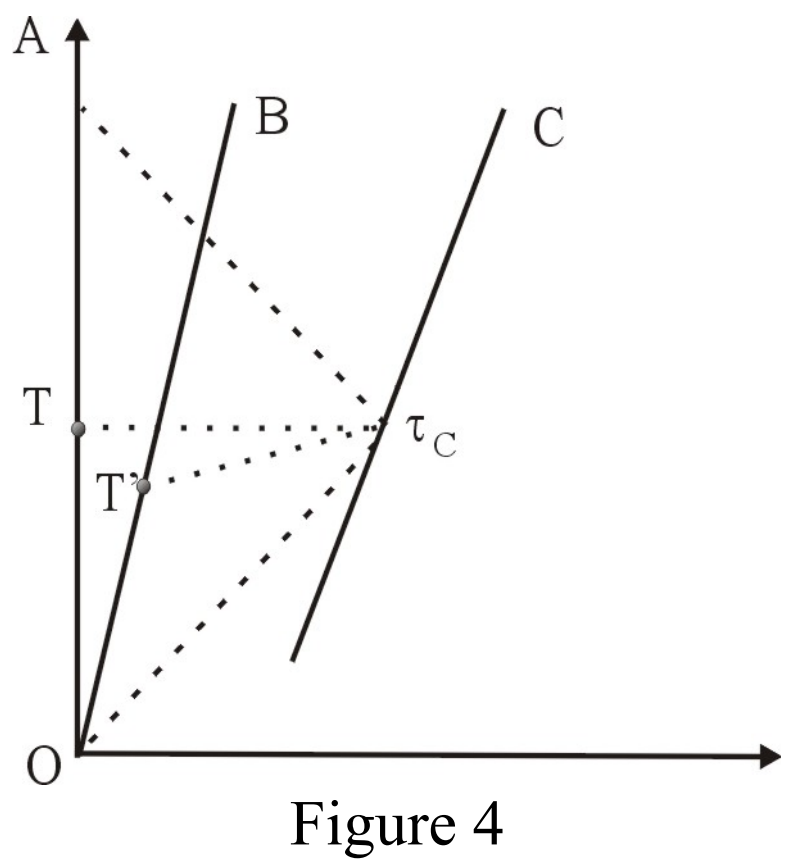

Figure 4

According to the clock of C the physical event "reflection of light by C" occurs at some value $\tau_C$ of her/his proper time. The phase of C' clock was not 'initially' set to that of A or B. She/he can take the moment at which this particular physical event occurs to set the phase to, e.g., zero, or simply leave the value as $\tau_C$. The moment at which light arrives at C can be characterized by the temporal value of her/his now-point. If fact from C's perspective her/his now-point is given by her/his position $x_C = 0$ and the value $\tau_C$ of her/his proper time. The trajectory of C in A and B's inertial reference frames, or more exactly her/his worldline can be seen as made of her/his successive now-points representing, e.g., the succession of moments on which C is reflecting light pulses send by A and B. Each now-point of C corresponds to a coordinate time of A and a coordinate time of B that do not coincide (as in the case depicted in figure 4). This is simply an example of the difference in the planes of simultaneity of two inertial observers in relative motion. If we now ask what is the 'set' of now-points (in terms of the values of the proper times of A, B and C) that are simultaneous, the situation becomes much more intricate than the simpler case of the two inertial observers with identical initial phase. In that case, as we have seen, when A reads, e.g., 10s she will consider that the simultaneous now-point of B corresponds to, e.g., t' = 8s; however this set of now-points is relative to adopting A as our reference observer, i.e. as the observer taken to be at (relative) rest. If we make the same question from B's perspective, when B reads 10s he, later, takes the corresponding now-point of A to be the one with the temporal value of t = 8s. In this case since they set their 'initial phases' to zero we can actually say that both have gone by 10s. When considering another inertial observer C that has not set her/his phase with that of A and B simultaneously we still can determine the set of simultaneous now-points relative to the adopted observer. In this way when A reads 10s, and takes B to read 8s, she considers that C corresponding moment is that in which her/his clock reads, e.g., $\tau_{C1}$. However B considers, when reading 10s, that A reads 8s and C reads, e.g., $\tau_{C1}$'. In this case we cannot determine what is the 'actual' value $\tau_C$ of the now-point of C that corresponds to A and B reading 10s since we lack, e.g., the initial simultaneous stipulation of the phase of C's clock with that of A and B.

This does not mean that there is nothing more we can say about the passage of time of C in relation to that of A and B. Let us consider the inertial motion of C (as determined by A and B) between her/his time reading $\tau_{C1}$ and $\tau_{C2}$. The amount of proper time gone by C as determined by A or B is given by $\Delta\tau = \tau_{C2} - \tau_{C1} = (t_2 - t_1)$ sqrt$(1 - (\upsilon_{CA}/c)^2)$= $(t_2' - t_1')$ sqrt$(1 - (\upsilon_{CB}/c)^2)$, where $t_1$ ($t_1$') and $t_2$ ($t_2$') are the time coordinates of the now-points corresponding to $\tau_{C1}$ and $\tau_{C2}$ as determined by A (B), and $\upsilon_{CA}$ ($\upsilon_{CB}$) is the velocity of C according to A (B). We again arrive at the time dilation formula. However, we already know that if an inertial body goes by a proper time interval of $\Delta\tau$, independently that other inertial bodies in relative motion ascribe to this time interval values determined by the time dilation formula $\Delta t = \Delta\tau/(1 - (\upsilon/c)^2)$, it goes by the same amount of proper time that these other inertial bodies go by/through. We arrive at the result that the inertial bodies are unfolding at the same pace given by the inertial time; something that we already knew from the principle of relativity.

This result can be seen as shared with classical mechanics, when reinterpreting Newton's notion of absolute time locally in terms of the proper time of material bodies (see, e.g., Misner et al. 1973, 289-90). However in Newton's case we take for granted that all inertial bodies have set their 'initial' time to the same value, even if at-a-distance (e.g. by an instantaneous action-at-a-distance; see, e.g. Brown 2005, 20; Torretti 1983, 13); and that all coordinate systems have identical time coordinates, according to each other (as we can check, e.g., from the Galilean/Galilei transformations; see, e.g., Torretti 1983, 28-9). This implies that the set of simultaneous now-points is clearly defined: (1) all inertial bodies unfold according to the inertial time; (2) all have their initial phase set to the same value; (3) the now-points of all is given/determined by the adopted coordinate time. Furthermore, conditions 1 to 3 apply also to the case of non-inertial bodies.

In the case of the theory of relativity, as we will see, we only have (1) for inertial bodies; (2) applies only to inertial bodies that have their temporal phase set initially in a way that we might consider simultaneous; (3) only applies in the case where (2) applies.

It might still be possible to think in terms of a three-dimensional spatial 'manifold' in which the inertial bodies co-exist. The temporal unfolding of all of them is given by the inertial time and this unfolding is 'fixed' to the three-dimensional 'platform'. However if we want to relate the proper time values of the now-points of all the inertial bodies this is made by resort to the exchange of light (that travels through the three-dimensional space between the inertial bodies). There is no unique way to 'sort' the temporal values of the now-points, since we have not set the phases of all of them as a sort of 'initial condition' (i.e. there is no unique set/net of simultaneous now-points established on the three-dimensional spatial 'manifold').[18]

To approach a more Newtonian outlook we might consider an extra hypothesis that of a sort of cosmic time (given by the shared inertial time with the same initial phase for all physical systems) for which conditions 1, 2, and 3 apply. For example, we impose a sort of big bang in which all material systems arise from a particular location at the same moment. However to do this in the context of the theory of relativity in which we use a four-dimensional manifold of events (independently of how we interpret it) seems to push too much the theory and not to do justice to the experimental situations in which the theory is applied, for which this kind of 'initial condition' seems out of place.

[18] It is beyond the scope of this work to try to articulate mathematically the idea of a three-dimensional spatial 'manifold' in which the unfolding of different physical systems might be given by their respective proper times, without having the initial phases of all the physical systems set 'simultaneously'.

The situation is less cumbersome than we might think when addressing the issue of the flow of time in the theory of relativity as compared to classical mechanics. The description of physical processes in the theory of relativity is made with inertial reference frames;[19] these inertial reference frames have their 'initial conditions' set: we take their initial event (0, 0) to coincide, in this way setting the phase of their master clocks. All of the master clocks give us the time lapse (which corresponds simply to the coordinate-inertial-proper time of each observer). When changing from one observer to another we must not get distracted by the time dilation and remember that all go through the same amount of inertial time. This means that if we change from observer A to observer B to describe physical events and processes and A's proper time reads 10s then we must 'pick up' B in his $10^{th}$ moment of time.

The temporal unfolding of all other material bodies and physical processes can be described in terms of this inertial time, marking the flow of time. Let us look at the case of non-inertial material bodies, i.e. material bodies that are accelerated. Let us consider an accelerated motion of a body D between two events in which D coincides with the observer C. The total amount of proper time gone by the body D is not the same as that of C; being accelerated the material body's proper time is given by the general Minkowski proper time integral $\Delta\tau_D = \int \text{sqrt}(1 - (\upsilon_{DA}(t)/c)^2)dt$ (here described in terms of A's coordinate system). In comparison to C the amount of proper time gone by D is smaller (see, e.g., Smith 1993, 49-55). However, and importantly for the discussion being made in this paper, both the now-point of C and the now-point of D are traced using the coordinate time of A (or B). This is possible because D co-exists with A, B, and C (e.g. it can be side-by-side with C and it exchanges light with A and B). However since we lack a further specification of an 'initial condition' regarding the relation of the phase of D with that of A and B we cannot answer a question like that of what is the 'actual' now-point of A and B when, e.g., C and D coincide and C reads $\tau_{C2}$ and D reads $\tau_{D2}$. When A receives back the light reflected by C and D she 'reconstructs' her past now-points that she considers to have been simultaneous to the now-points of C and D corresponding to their time readings of $\tau_{C2}$ and $\tau_{D2}$. If B does the same procedure he will arrive at a different time coordinate.

Again we face the situation that different inertial observers will consider a different net/set of now-points to be simultaneous, and it does not seem to be possible in general to find the 'actual' set of simultaneous now-points. For this it seems necessary a further specification of 'initial conditions'. This occurs in the very particular case in which the accelerated body D is side-by-side with A and B at the beginning of the 'experiment' and has its phase set to zero simultaneously with A and B. In this case we can answer the question. A and B trace the motion of D, e.g., by emitting light pulses to D that are reflected back. In this way A and B determine the relation between the proper time of successive now-points of D in relation to their respective coordinate time. This is done, as mentioned, by 'reconstruction': each of them determines her/his past value of the proper time that is on her/his simultaneity plane passing by the worldline of D, corresponding to its proper time $\tau_D$. But let us now consider that, e.g., the now-point of D corresponding its proper time $\tau_D$ corresponds also to the now-point of a hypothetical inertial observer C (see figure 5). We imagine C to have her/his phase set to zero simultaneously with A, B, and D. Let us say that the proper time along C's worldline from the origin to the event corresponding to D's reading of $\tau_D$ is equal to $\tau_C$. This implies that the now-point of D corresponds to a passage of inertial time equal to $\tau_C$, i.e. we can say that D's now-point has an inertial temporal value of $\tau_C$. The hypothetical observer C is taken to be physically equivalent to A or B. This means that if C goes through $\tau_C$ seconds, then also A and B have gone by this same amount of time (independently of what, due to the time dilation, each of them considers to be the simultaneous now-point of the other). This sets in a unique way the net/set of simultaneous now-points independently of the observers, as in classical mechanics.

[19] Non-inertial reference frames can be used, but their applicability is limited. In simple terms they cannot cover all the four dimensional space of events (see, e.g. Callahan 2000, 143-165)

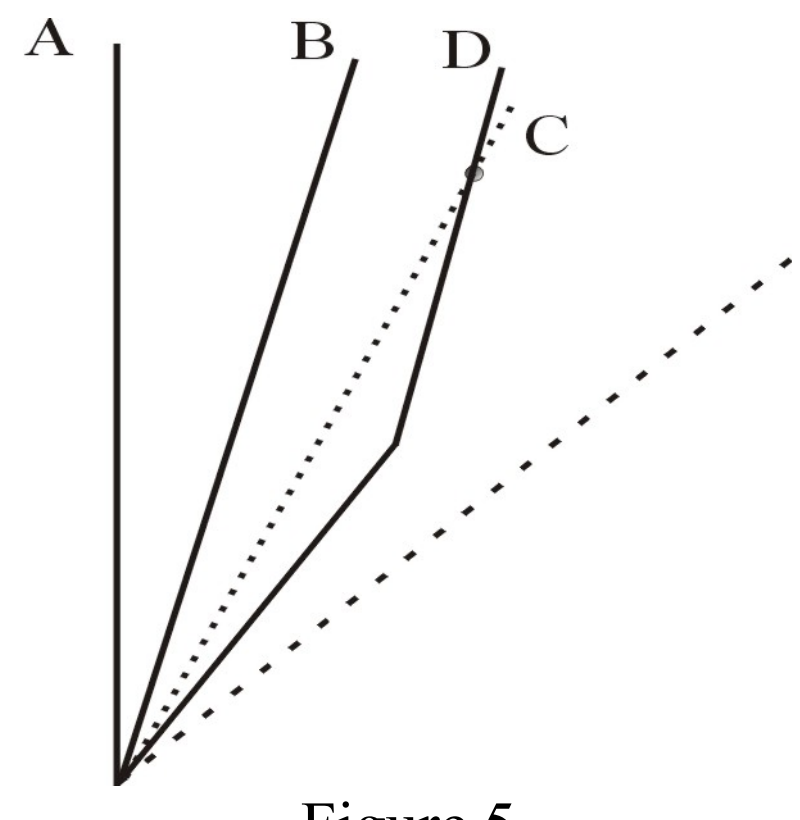

Figure 5

In this quasi-Newtonian situation, conditions 1 to 3 apply to the inertial bodies in question (A, B, and C). The proper time of the non-inertial body D unfolds differently from the inertial time; however there is a one-to-one correspondence between the 'actual' now-point of D and the 'actual' (identical) now-points of A, B, and C. Since the proper time of D is determined as a function of the inertial time of one of the inertial observers, one can consider that its unfolding is a function of the inertial time and its now-point is determined also by the inertial time. This means that for this particular case, with some further clarification of the relation of the accelerated body's proper time to the inertial time, we see that conditions 1-3 apply also to the case of the accelerated body.

## 6 Epilogue: the relativity of simultaneity

According to the previous section if we choose, e.g., two inertial observers A and B whose coordinate systems are defined in a related way by setting their respective initial phase and origin as the same event, the relation between the now-points of these observers is clear: the 'actual' temporal value of the now-points of the two observers is the same even if due to the time dilation they might consider that the other observer corresponding now-point is one with a smaller temporal value; i.e. if A's now-point is the one with $\tau_A = 10s$ then B's now-point is the one with $\tau_B = 10s$.

Even if each observer determines a different net/set of simultaneous now-points (corresponding to a different determination of the simultaneity plane), they can nevertheless know, by resort to the principle of relativity, that both have now-points with the same temporal value.

In general, this unique temporal value of the now-points cannot be 'spread' to other inertial bodies. The now-points of these are constrained by their spatial co-existence with A and B and by the fact that they also unfold according to the inertial time; how-ever it does not seem to be possible to determine a unique set of simultaneous now-points. For this it seems necessary a further specification of an ‘initial condition’ (e.g., by a simultaneous setting of the time phase of all material bodies under consideration). Only in this case we approach a more Newtonian situation in which all the now-points of the inertial bodies have the same temporal value.

The case of non-inertial bodies brings more complexity to the issue. Again we face the situation that in general it does not seem possible to determine a unique net/set of simultaneous now-points; furthermore the unfolding of an accelerated body is not the same as that of an inertial body. However, as remarked, this unfolding can be seen as a function of the inertial time. For the particular case in which an accelerated body has its phase set to that of the inertial observers A and B we can in fact relate its unfolding now-point in a unique way to the unfolding now-points of A and B (given by the same temporal value).

Even if with limitations and a clear difference to the Newtonian case it seems that there is a place for a notion of flow of time in the theory of relativity along the lines of Dieks’ proposition.

Having settled in this ‘provisionary’ state of affairs regarding the flow of time in the theory of relativity, let us return to the issue of the relativity of simultaneity in relation to the flow of time. Let us consider the special situation in which four inertial observers in relative motion set their phases to zero when initially side-by-side (see figure 6).

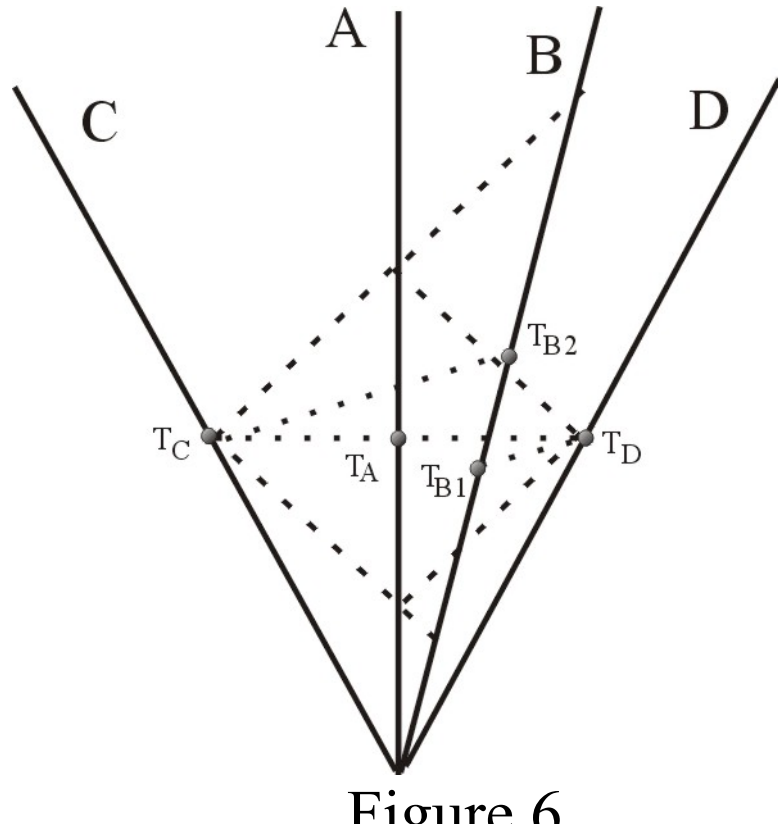

Figure 6

As it is well known, observer A will take the events along the worldlines of C and D corresponding, e.g., to $T_C = T_D = 10s$ to be simultaneous, having in A’s coordinate system, e.g., the value $T_A = 12s$; due to the time dilation even if A takes the now-points of C and D to be simultaneous she attributes to them a value greater than their proper time readings. In the case of observer B he considers that the events along the worldlines of C and D corresponding to $T_C = T_D = 10s$ occur at the different moments $T_{B1}$ and $T_{B2}$. Regarding the relation between the temporal values of the now-points of C and D and A and B, both A and B are in a way wrong! When C and D go through 10 seconds, A and B also go through 10 seconds.

In general since we do not have the apparently necessary initial condition it seems impossible to determine, besides the ‘actual’ identical temporal value of the now-points of A and B, the ‘actual’ temporal values of the now-points of other material bodies. Different observers consider a different net/set of past now-points to have been simultaneous.

As we have seen, this situation does not need to be seen as an impediment to a local view of the flow of time, for in this case it is unnecessary, as Dieks remarks, to have “a succession of a definite set of global simultaneity hyperplanes in order to accommodate our experience” (Dieks 2006, 160). In Dieks’ view only along each worldline is there a clear order relation between successive now-points. Accordingly, “this complete linear order of now-points is the theoretical representation that fits the Newtonian conception of time” (Dieks 1988, 458). When we consider the relation between the now-points of different worldlines (i.e. the now-points of different material bodies) their relation is constrained by the linear order of the successive now-points along each worldline (arising from the inertial time) and the spatial co-existence of physical systems and processes; but in general, as we have seen, this does not result in a unique relation (there are different sets of now-points taken to be simultaneous depending on the observer). Dieks refers to this situation as that of having only a partial order in the relation of successive now-points of different worldlines (Dieks 1988, 459; Dieks 2006, 171). However as we also have seen there are cases in which we might consider that this partial order is a linear order falling into the canon of Newton’s conception of time. For this to be possible we need to consider, e.g., the special case in which different material bodies (inertial an eventually also non-inertial) are ‘initially’ side-by-side and have their respective temporal/time phase set to the same value.

The relation between the now-points of different inertial observers (related by the Lorentz transformations) corresponds to this particular case. This means that for all observers involved in the description of physical events and processes we can keep track of their now-points: all have the same temporal value as given by the coordinate time.

It is an open issue if we can take more seriously the idea of a three-dimensional spatial ‘manifold’ in which each material body unfolds according to its proper time. For particular cases where, e.g., several particles, some inertial others not, ‘departure’ from the same point at the same time this image seems simple to maintain; for the general case it is not clear. However we must bear in mind that even the putative general situation presented in terms of the four-dimensional Minkowski space of events has inbuilt a particular choice of ‘initial conditions’. One considers the Minkowski space-time coordinatized in terms of two inertial coordinate systems corresponding to two inertial reference frames that have a coinciding space and time origin. All isolated events or time-like and light-like worldlines are described at least with this minimum initial condition, even if in general the time phases of different inertial and non-inertial worldlines are not set in a non-ambiguous way (i.e. in way in which the relation between the different now-points taken to be simultaneous does not depend on the chosen coordinate system).

## Appendix: interpreting the Minkowski space-time diagrams

Minkowski’s space-time approach to the theory of relativity was made with resort to a particular type of diagrams – the so-called Minkowski space-time diagrams –, which due to their widespread use became influential in the interpretation of the theory, in particular in relation to the issue that occupies us here, the eventual flow of time.

In its most basic form, a space-time diagram simply depicts the space-time as coordinatized with a particular inertial coordinate system, which means that we represent a time axis and usually just one spatial axis, e.g. an x-axis. Then, in relation to this adopted coordinate system, we also represent the set of events corresponding to the t’-axis and the x’-axis of another inertial coordinate system. The depiction of the axes is complemented, as necessary, with the representation of, e.g., invariant sets of events like the light cone or hyperboles given by $t^2 - x^2 = k$ (with $k > 0$ or $k < 0$); or one simply represents, e.g., worldlines of light signals/pulses and worldlines of inertial or non-inertial material bodies.

In Minkowski's original presentation, the space-time diagram was used to provide a 'graphical' deduction of the Lorentz transformations between inertial coordinate systems, and a 'visual' presentation of the so-called Lorentz contraction (Minkowski 1908; Ohanian 2012, 17; Walter 2010, 11-4). In fact, the space-time diagrams provide a very direct depiction of all the relativistic effects. In this work we used space-time diagrams in the discussion of the time dilation and the relativity of simultaneity.
There is a view that space-time diagrams express the eternalist block structure of the Minkowski space-time (see, e.g, Petkov 2005). This precludes any physical notion of flow of time. It is evident that this view is incompatible with Dieks' view in terms of a now-point per worldline. According to Dieks,

> the four-dimensional spacetime diagram *records* events with their qualities and relations. But in order to be recordable at all, the events in question must *occur* ... non-occurring events are evidently not represented in the four-dimensional picture. (Dieks 2006, 170)

In simple terms we may say that only measured/observed events (i.e. events that have become) are represented in a space-time diagram. Regarding the flow of time and becoming, Dieks concludes that "nothing has to be added to the space-time diagram: the four-dimensional picture *already contains* becoming" (Dieks 2006, 174). In this way the Minkowski diagram represents events that have occurred and have been measured, and the flow of time is, in a way, 'inscribed' in the diagrams.

Let us look in more detail into how Dieks 'extracts' the now-points and their relations from the space-time diagrams. Dieks starts by considering just a space-time diagram representing one worldline. According to Dieks "the idea of flow of time in this case can be implemented by singling out 'now-points' on the worldline" (Dieks 1988, 458). By assigning successive now-points to the worldline, this leads to "an infinitude of space-time diagrams, individuated by different locations of the now-point" (Dieks 1988, 458). Here we have the case Dieks' refers to as a Newtonian-like linear-order of now-points (of the same material body).

If we have two or more worldlines represented in a space-time diagram the situation is more complex. We can assign (successive) now-points to each worldline submitted to the restriction that they are representing physical systems that spatially co-exist. This implies that no now-point is located in the conjunction of the past light cones of the other now-points. Like in the simpler case of just one worldline we "obtain an infinitude of space-time diagrams each distinguished by its own assignment of now-points" (Dieks 1988, 459). However in this case

> it is not possible to order these diagrams according to the order of the linear continuum. Instead, our collection of space-time diagrams is a set that is only *partially ordered*. The partial order is generated by the usual linear order along worldlines: there is in general no order relation between diagrams that are generated from one original diagram by moving the now-points along two different worldlines. (Dieks 1988, 459)

This does not mean that in any diagram or 'sequence' of diagrams in which one explicitly represents now-points it is not 'inscribed' the flow of time: the diagrams represent events that have occurred or, less strictly, events that we might expect to occur in a particular experiment. Also, as mentioned, there are particular cases in which we can ex-plicitly represent simultaneously in one space-time diagram (or a sequence of diagrams) the actual now-points along several worldlines, as already done in this paper.

Another proponent of the view that proper time gives the flow of time is Arthur (2006, 2008). His view on the space-time diagrams is very similar to Dieks', even if formulated in part implicitly. According to Arthur, "once we have represented all events and all processes on a space-time diagram, we have represented all becoming" (Arthur 2006, 135-6). It is evident that in Arthur's view the diagram is a representation of past occurring events and processes, which we must consider to have been measured/observed (or expected-future present and past occurring events). Arthur is more explicit in another remark:

> in constructing a spacetime diagram we represent processes and events, that is, things that are supposed to have occurred. Becoming – or at least, having become – is already included in the diagram. (Arthur 2006, 151)

There would be no contradiction between the implementation of the theory of relativity in terms of the Minkowski space-time using Minkowski diagrams and the view that the theory of relativity enables a notion of flow of time.

This interpretation of the Minkowski diagrams is not completely new. In fact it can be found, at least in part, in Bohm's introductory book on the theory of relativity, even if Bohm does not address explicitly the flow of time (Bohm 1965). In chapter 31, entitled "the significance of the Minkowski diagram as a reconstruction of the past", Bohm presents a view on the Minkowski diagrams that is similar to Dieks' and Arthur's views. Bohm asks us to consider an 'observer' at rest in a laboratory (taken to be an inertial reference frame). In a Minkowski diagram we represent the observer as a straight line OA (see figure 7). Let us consider that for the observer 'now' corresponds to the event P.

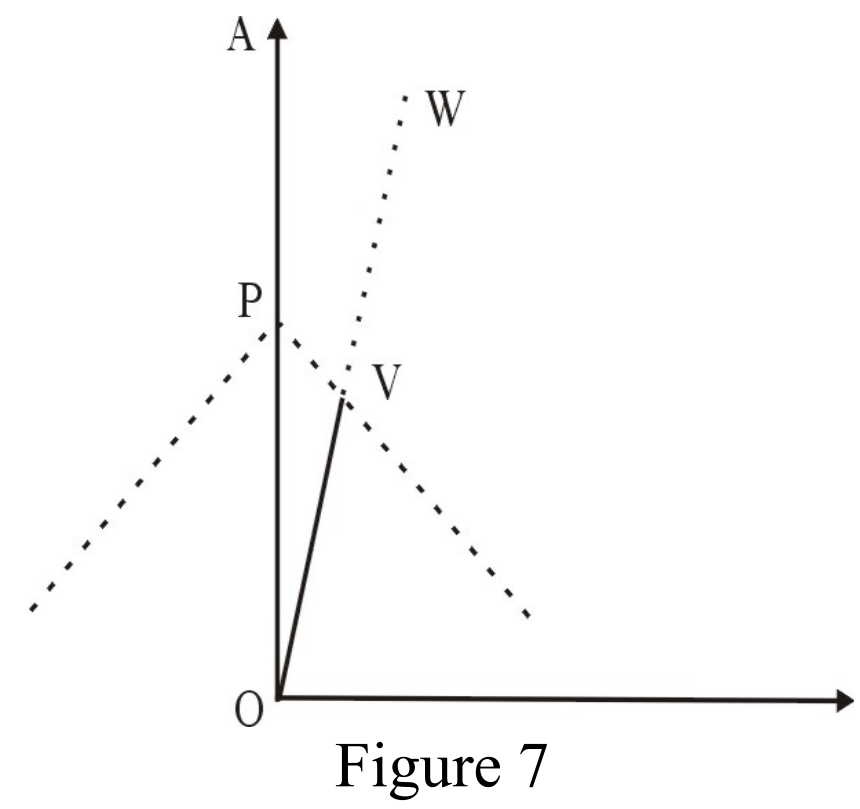

Figure 7

According to Bohm, the observer "cannot survey the whole Minkowski diagram. On the contrary, he can only know of events that are inside his past light cone" (Bohm 1965, 174). If we imagine that the observer is measuring/determining the worldline of a particle, when 'now' at P the observer has information regarding the particle only up to event V. According to Bohm, the observer

> can reasonably assume that the particle continues to exist on the line VW, which is the extension of OV into the region outside the light cone of P ... but this projected picture is always subjected to contingencies. (Bohm 1965, 176)

The Bohm-Minkowski diagram can be seen as a reconstruction in a mathematical space representing events (i.e. length and duration measurements) of past measurements or expected-future present and past events, like the worldline VW (resulting from future measurements/observations). In this way it includes, even if implicitly, a reference to now-points and the flow of time. For example, in the Bohm-Minkowski diagram of figure 7 it is represent: (1) the now-point P of an observer; (2) past events inside her/his past light cone, like the (past) worldline OV; (3) expected-future present and past events, like the future now-points of A (implicit in the vertical axis) and the worldline VW.

This view makes it necessary to be careful in the interpretation of measurement results (or theoretical predictions) as represented in a diagram, in particular if one forgets about how the diagram is obtained and that in its construction is implicit the flow of time and the notion of now. In the diagram we can represented not yet observed expected future events, but we do not have to conclude from this that the Minkowski space-time and the Minkowski diagram imply a block universe in which all past and future are given all at once, as Dieks (2006) and Arthur (2006) call our attention to.